\documentclass[11pt]{llncs}

\usepackage{listings}
\usepackage{xspace}

\title{The MathScheme Library: Some Preliminary
  Experiments\thanks{This research was supported by NSERC.}}

\author{Jacques Carette \and William M. Farmer \and Filip Jeremic \and
  Vincent Maccio \and Russell O'Connor \and Quang
  M. Tran\thanks{\texttt{\{carette,wmfarmer,jeremif,tranqm\}@mcmaster.ca},
    \texttt{vincentmaccio@gmail.com}, \texttt{roconnor@theorem.ca}.}}

\institute{%
Department of Computing and Software\\
McMaster University\\
Hamilton, Ontario, Canada
\\[1.5ex]
13 May 2011 \vspace*{-4.5ex}
}

\begin{document}
\maketitle

\lstdefinelanguage{mathscheme}
    {morekeywords={Theory,combine,extended,by,type,over,axiom,
        implies,not,forall,and,not,or,Inductive,case,of,instance,via,
        defined-in},
    basicstyle={\small},
    keywordstyle=\bfseries}

\lstset{language=mathscheme}

\lstMakeShortInline|

\begin{abstract}
We present some of the experiments we have performed to best test
our design for a library for MathScheme, the mechanized mathematics
software system we are building.  We wish for our library design to
use and reflect, as much as possible, the mathematical structure present
in the objects which populate the library.
\end{abstract}

\section{Introduction}

The mission of mechanized mathematics is to develop software systems
that support the process people use to create, explore, connect, and
apply mathematics.  The objective of the MathScheme
project~\cite{mathscheme} is to develop a new approach to mechanized
mathematics in which computer theorem proving and computer algebra are
merged at the lowest level.  Our short-term ($2$--$3$ years) goal is to develop
a framework, with supporting techniques and tools, for tightly
integrating formal deduction and symbolic computation.  The long-term
($7$--$10$ years) goal is build a mechanized mathematics system based on this
framework.

A critical component of any mechanized mathematics system is a large library of
formalized mathematics.  We believe such a library should be constructed from
modular units representing theories and theory morphisms.  The library should
be equipped with powerful methods for building complex knowledge structures by
combining and relating theories and theory morphisms.  It should include
mathematical knowledge expressed both declaratively (using axioms) and
constructively (using algorithms).  And it should equally serve users
who want to explore and apply the knowledge in the library and
developers who want to organize and expand the knowledge in the
library.

The design process for the MathScheme Library is being driven by a
number of key ideas motivated by lessons learned from previous
endeavors (ours as well as other's):

\begin{enumerate}

  \item \emph{Abstract Theories}.  Abstract axiomatic theories, such
    as the familiar theories of abstract algebra, can be highly
    interrelated and, as a result, an unsophisticated formalization of
    abstract axiomatic theories will become bloated with redundancy as
    it grows larger.  Can the \emph{tiny theories method}, in which
    mathematical knowledge is organized as a network of theories built
    up one concept at a time, be used to systematically
    eliminate the harmful forms of this redundancy?

  \item \emph{Concrete Theories}. A concrete theory, as opposed to an
    abstract axiomatic theory, is a description of a specific
    mathematical structure and often serves as a basis for
    computation.  Can concrete theories be developed using the same
    techniques as for developing abstract axiomatic theories?

  \item \emph{Applied Universal Algebra}.  Universal algebra includes
    many useful algebraic constructions that can be applied to a wide
    variety of mathematical structures.  Can these constructions be formalized
    uniformly as operations on theories?

  \item \emph{Biform Theories}.  A \emph{biform theory}~\cite{Farmer07b} is a
    combination of an axiomatic theory and an algorithmic theory.  Can
    biform theories uniformly replace axiomatic theories in libraries
    of formalized mathematics?

  \item \emph{Theory Implementations and Interfaces}.  An
    \emph{implementation} is a theory whose concepts and facts are
    divided into primitive and derived, while an \emph{interface} to an
    implementation is a theory that contains some of the concepts and
    facts of the implementation but ignoring their status with respect
    to being primitive or derived.  Can a library of theories be
    organized so that some theories are interfaces to several
    implementations and every implementation has one or more
    interfaces?

  \item \emph{High-Level Theories}.  A \emph{high-level
    theory}~\cite{CaretteFarmer08} is a high-level environment for reasoning and
    computation that is analogous to a high-level programming
    language.  Can high-level theories be built on top of a networks
    of abstract axiomatic theories and concrete theories?

\end{enumerate}

Although these key ideas seem to be sound in theory, they have not
been sufficiently tested in practice.  In particular, it is not
exactly clear how they should be implemented to meet the requirements
of a ``real'' system.  For this reason we are conducting a series of
design-and-implement experiments to learn how to best incorporate
these ideas into the MathScheme Library.  We are particularly
interested in exploring the impact these ideas can have on the
\emph{scalability} of contemporary libraries of formalized
mathematics.  In this work-in-progress paper, we report on experiments
dealing with the first three ideas: abstract theories, concrete
theories, and applied universal algebra.

We are working on making the details of these experiments available; details
will soon appear on the project's web site~\cite{mathscheme}.

\newcommand{\abbr}[1]{\texttt{\small #1}\xspace}
\newcommand{\U}{\abbr{U}}
\newcommand{\Carrier}{\abbr{Carrier}}
\newcommand{\Pointed}{\abbr{Pointed}}
\newcommand{\Magma}{\abbr{Magma}}
\newcommand{\PointedMagma}{\abbr{PointedMagma}}
\newcommand{\mult}{\abbr{*}}
\newcommand{\e}{\abbr{e}}
\newcommand{\bit}{\abbr{bit}}

\section{Abstract Theories}\label{sec:abstract}

As is quite well-known, the axiomatic theories of \emph{Algebra} are highly
interrelated.  Theories in other areas of mathematics also seem to be 
quite structured, but they have been subject to less intense classification
work.  The question is, what is needed to build up a sufficiently
rich library of algebraic theories, while minimizing the human labor needed
to create and, at the same time, maximize sharing between theories?

For example, we know that a Field is a commutative Ring which is also a
Division Ring.  Similarly, a Ring combines a non-unital ring (often called a
Rng) with a SemiRing.  Can these relations be used in the explicit 
construction of an algebra hierarchy, in a way which is both semantically
meaningful as well as labor-saving?  We strongly believe that the
answer is a resounding ``yes''.  

\subsection{The ideas}

We needed to test whether the relations between axiomatic theories could be 
leveraged in a useful way in building a (large) library of mathematics.
We needed to understand exactly \emph{which} relations could be used
(rather than being shown to exist) in the building of the library.

We reasoned that the most important relation, even though it is in
fact the simplest, is that of \emph{inclusion} at the level of
\emph{theory presentations}.  In other words, even though we are
(eventually) interested in the semantics of theories, it is at the
level of the \emph{syntax} where we can gain the most, at least from
the point of view of building up a library of abstract theories.

A second idea to test is that of \emph{tiny theories}: each separate concept
should occur once and only once in the library \emph{source code}, even
though the semantic concept may well be pervasive.  For example, the
concept of a binary operation being |commutative| should occur only once.

\subsection{The experiment}

To pick up on the example of Field and Ring, our library source contains the
statements
\begin{lstlisting}
Ring := combine Rng, SemiRing over Semirng 
Field := combine DivisionRing, CommutativeRing over Ring
\end{lstlisting}
\noindent defining Ring and Field respectively.\footnote{The semantics will be
described later on, we hope the syntax is sufficiently evocative to not need a
detailed explanation at this moment.}  But where does a Ring structure really
come from?  More precisely, where do a (multiplicative) semigroup and an
(additive) monoid first ``cross'' to form the core of a ring?  By combing
through sufficiently detailed algebra textbooks, one encounters the
notion of a \emph{left near semiring} which seems to fit the bill.  And
indeed, we define
\begin{lstlisting}
LeftNearSemiringoid := combine Semigroup, AdditiveMonoid 
                         over Carrier
LeftNearSemiring := LeftNearSemiringoid extended by {
  axiom leftDistributive_*_+ : leftDistributive((*),(+));
  axiom left0 : leftAnnihilative((*),0)
}
\end{lstlisting}
\noindent where the |LeftNearSemiringoid|\footnote{This is our name for this
structure; we could not find another name in the literature.} is a pure
combination of a |Semigroup| and an |AdditiveMonoid| which share the same
|Carrier| set and nothing else.  A |LeftNearSemiring| then adds two axioms,
left distributivity of |*| over |+| and that |0| is a left annihilator for |*|.
Note how this second definition does not in fact properly obey the rules
of \emph{tiny theories}: it introduces two concepts at once.  We have thus
discovered that there are in fact two intermediate algebraic structures
in between a |LeftNearSemiringoid| and a |LeftNearSemiring|.

\begin{figure}[th]
\begin{lstlisting}
LeftNearSemiring := Theory { 
  U : type;
  * : (U, U) -> U;
  + : (U, U) -> U;
  0 : U;
  axiom right_Identity_+_0 := forall x : U.  x + 0 = x;
  axiom leftIdentity_+_0 := forall x : U. 0 + x = x;
  axiom leftDistributive_*_+ := 
    forall x,y,z : U. x * (y + z) = (x * y) + (x * z);
  axiom left0 := forall x : U. 0 * x = 0;
  axiom associative_+ := 
    forall x,y,z : U.  (x + y) + z = x + (y + z);
  axiom associative_* := 
    forall x,y,z : U.  (x * y) * z = x * (y * z)} 
\end{lstlisting}
\caption{LeftNearSemiring theory presentation}
\label{fig:lnsr}
\end{figure}

But what is a |LeftNearSemiring|?  Figure~\ref{fig:lnsr} shows a more
``classical'' presentation of this theory.  It shows a carrier type
\U, a constant |0|, and two operations |*| and |+| over that carrier
type, and the six axioms of a |LeftNearSemiring|.  We call the theory
in Figure~\ref{fig:lnsr} the \emph{expanded} version of the
|LeftNearSemiring| theory (defined above).  This expanded version is
what we are interested in specifying, but not what we want to enter
(as a human): this would be incredibly tiresome as well as error-prone
to do for even a small number of theories.  We are flabbergasted
that this is nevertheless essentially how it is done in all current
large libraries of mathematics, either for theorem proving purposes or
for computer algebra.

\begin{figure}[bht]
\begin{lstlisting}
Empty := Theory {}
Carrier := Empty extended by { U : type }
PointedCarrier := Carrier extended by { e : U }
UnaryOperation := Carrier extended by { prime : U -> U }
PointedUnarySystem := 
  combine UnaryOperation, PointedCarrier over Carrier
DoublyPointed := PointedCarrier extended by { e2 : U }
BinaryOperation := Carrier extended by { ** : (U,U) -> U }
Magma := BinaryOperation [** |-> *]
CarrierS := Carrier[U |-> S]
MultiCarrier := combine Carrier, CarrierS over Empty
UnaryRelation := Carrier extended by { R : U ?}
BinaryRelation := Carrier extended by { R : (U,U)?  }
InvolutiveUnarySystem := UnaryOperation extended by { 
  axiom involutive_prime : 
    forall x:domain(prime). prime(prime x) = x
}
Semigroup := Magma extended by { 
  axiom associative_* : associative((*)) }
\end{lstlisting}
\caption{Base of theory hierarchy}
\label{fig:base}
\end{figure}

Significantly more difficult was building the ``base'' of this 
theory\footnote{Strictly speaking, these are all theory \emph{presentations}
rather than theories, but we will address this point later.}
hierarchy.  Figure~\ref{fig:base} shows the first few lines.  The reason
this was difficult was that it took a certain amount of time to convince
ourselves that we really needed that many ``trivial'' theories in order
to later reap the benefits of this extreme modularization.

At the root of this network is the empty theory containing no concepts
(e.g., types or operations), although the full \emph{internal} logic
of the system is implicitly present in the |Empty| theory.  A theory
called \Carrier extends the empty theory adding a universe type \U.
Theories can be extended in multiple ways.  One extension of \Carrier
called \Pointed adds a constant \e of type \U.  Another extension of
\Carrier called \Magma adds a binary function \mult of type |(U,U) -> U|, 
which it ``gets'' from |BinaryOperation| through a renaming.
Other extensions adds axioms to the theory.  For example, the theory
called |Semigroup| adds to \Magma an associative axiom for \mult.
Each extension defines an inclusion at the level of theory
presentations, from the smaller theory into the larger.  At the level
of the theories themselves, the induced theory morphisms are
considerably more complex.

Using this process of building algebraic theories, we have built up $201$
(presentations of) theories, including |BooleanAlgebra|, |Diod|,
|KleeneAlgebra|, |MoufangLoop|, |OrthomodularLattice|, |Quandle|,
|StarSemiring| and |VectorSpace|, and a large network of morphisms relating
these theories to each other.

\subsection{The method}\label{sec:abstract:method}

We have a formal grammar for the MathScheme language, as well as a
notion of how certain terms \emph{expand}.  The underlying semantics
is based on the category of \emph{presentations} of multi-sorted
theories (over the logic Chiron~\cite{Farmer07a,Farmer11}), where
theory morphisms are induced by signature mappings.  In other words,
for theories $T$ and $S$, $T\rightarrow S$ if there is a renaming
$\rho$ of $T$ such that $\rho(T) \subset S$, where inclusion is with
respect to intensional equality of all components of a theory.
|extended by| and renaming (see the definition of |Magma| in
Figure~\ref{fig:base}) induce the obvious morphisms. A statement like
|combine A,B over T| then means the \emph{pushout} of the induced
diagrams, where it is assumed that we can infer the morphism from |T|
to |A|, and |T| to |B|.

We have code which implements these constructions.  More precisely, we
have a base language of theories (very classical) as well as theory
combinators.  These combinators are ``theory constructions'', which
can be ``expanded'' according to the (informal) semantics above.
Figure~\ref{fig:lnsr} shows the actual result of expanding
|LeftNearSemiring|.

\subsection{The results}

This particular experiment is our most successful.  Both ideas (that it is
at the level of the syntax of theory presentations where there is the most
reuse, and that to achieve this every concept should be presented only
once) really have proven themselves.  What we did discover, however, is
that we focused too much on \emph{theories}, and that the most important 
structure is really present in the \emph{theory morphisms} induced by
our constructions.  We are currently conducting further experiments based
on this new knowledge.

It is important to note that our claims to novelty (if any) are largely
on the engineering aspects of our library: while the 
underlying ideas are old, we have not found anyone who has pursued these 
ideas as systematically as we have, nor to the scale which we have.  A
significant part of our infrastructure work was forced upon us because
of the scope of our library.

\section{Concrete Theories}\label{sec:concrete}

A concrete theory is often known as a \emph{structure}.  While most abstract
theories admit many models (even up to isomorphism), concrete theories are
those which \emph{by construction} admit a single model (again, up to
isomorphism).

\subsection{The ideas}

Basically we wanted to see if the same ideas that we used for abstract
theories also worked for concrete theories.  Certainly many concrete
theories are well-known to be \emph{parametric}, but what other
structure could we leverage?

\subsection{The experiment}

While the most fundamental concrete theories are the empty theory and
the |Unit| theory (of a singleton carrier set), the first important,
non-trivial concrete theory has multiple names: $2$, \bit, and
|bool|.  One of the simplest presentations of this theory can be given
in our language as
\begin{lstlisting}
Bit_Base := Empty extended by {
  Inductive bit 
    | 0:bit
    | 1:bit
}
\end{lstlisting}
\noindent It should be noted that 0 and 1 here are simply identifiers, and
carry no special meaning to the system.  It is also possible to give an
axiomatic presentation of the same theory, viz.
\begin{lstlisting}
Bit_Base_Abstract := Empty extended by {
  bit : type 
  1 : bit
  0 : bit
  axiom: forall b:bit. b = 1 or b = 0
  axiom: not(1=0)
}
\end{lstlisting}
\noindent which is isomorphic to the previous theory.  We prefer the first
(more functional) presentation as we get the axioms ``for free'' by the
definition of an inductive type in the underlying logic. 

Here too we try to follow the same principle as before, which is to try
to augment each of our (concrete) theories with a single concept at a time.
This also makes reuse much simpler.  For example, we may wish to define a
concrete function |and| between |bit|s.  (It is named |bit_and| so as to not
clash with the |and| from the internal logic).
\begin{lstlisting}
Bit_And := Bit_Base extended by {
  bit_and : (bit, bit) -> bit;
  bit_and(x,y) = case x of {
    | 0 -> 0
    | 1 -> y
  }
}
\end{lstlisting}
\noindent As can be seen, the language provides pattern-matching for
inductive types.  A more comfortable theory of |bit|s would combine more
operations, as indicated by our ``basic'' |Bit| theory:
\begin{lstlisting}
Bit := combine Bit_And, Bit_Or, Bit_Not, 
               Bit_Implies, Bit_Xor, Bit_Xnor over Bit_Base
\end{lstlisting}
\noindent These pieces, via renaming, augmented with additional operators
(like modal operators) can also be used for creating various logics.

We can proceed in the same manner for a theory for characters, and similar
enumerative theories.  In essentially the same way, we can also define
the natural numbers, following the classical definition of Peano\footnote{%
Naturally these natural numbers will only be used in proofs and properties.
We will need a better representation for actual computations.}.

What seems next, at least for computer scientists, would be a theory
of finite sequences of bits (words).  Experience
tells us that the proper way to do this first involves creating
polymorphic theories for (finite) sequences.  A sequence is just a (total)
function from |nat| (seen as a countable linear order) to a set.  
A finite sequence can be modeled in at least four different ways: as
a restriction of a sequence to an initial segment (of |nat|), as a 
\emph{partial} sequence guaranteed total on an initial segment, as a (total)
function on an initial segment, or as a list of elements.  

In other words, a sequence is not quite a concrete theory, as the above models
are not entirely equivalent.  We can get one concrete version by specializing
the theory of lists to bits:
\begin{lstlisting}
List := Carrier extended by {
  Inductive list
    nil : list
    cons : U -> list -> list;
}
BitCarrier := instance Bit_Base of Carrier via [ bit |-> U ]
BitList := combine List, BitCarrier over Carrier
\end{lstlisting}
\noindent  An |instance| encodes a non-inferable arrow, which is needed to
make the |combine| (pushout) work properly.

We could similarly instantiate the other models, and
each has advantages and disadvantages.  Ultimately, all four should be
available (and proven equivalent), but at this point we needed to make a
choice.

In reality, we would not form the |BitList| theory as an instance of just
the carrier, but rather from a (conservative) extension with convenience
functions like |length|, |map|, |zipWith|, etc.\ added.  Similarly, we would
really want to combine an enriched |List| with an enriched |Bit| theory
to form a ``useful'' theory of finite bit strings.

We have developed theories for a variety of data-structures (trees,
graphs, lists, stack, queue, dequeue, multiset, functional maps, etc.), 
various kinds of numbers, some machine-oriented types,
as well as some algorithms over these.  For example, we have a model
of the SHA 256 algorithm specified (constructively).

\subsection{The method}

The work was done in a different manner: rather than immediately start 
with \emph{tiny theories}, we started with more classical axiomatizations
(as found in a variety of textbooks) for the various structures.  The 
hope was that we could then ``see'' the relations between structures, and
perform stepwise abstractions from our (large) theories into a network of tiny
theories.

We also started out by writing a lot of the theories in a rather relational
(axiomatic) style, and only realized part way through that most of these also
admitted purely functional, fully constructive axiomatizations.  As these
are easier to leverage, as well as being more appropriate for concrete
theories, some rewriting was necessary.

\subsection{The results}

While we have been working on building concrete theories for two years,
it is safest to call this experiment as being fully in-progress.  We have
rewritten most of our library of concrete theories twice now --- and will
likely do so again.  We keep finding new ways to express these theories
which nicely factor out common components.  However, the kinds of
commonalities we find seem to be of a somewhat different kind than that
present in abstract theories.  Thus we need different tools to capture
these relations.

Particularly intriguing is that sometimes the very same semantics (classical
constructions in category theory, in particular colimits) is best specialized
into a number of different features, rendered with quite different syntax.
Concrete theories are ``instances'' of abstract theories, and we are still
trying to fully leverage the consequences of this.

We have definitely learned that concrete theories need to be defined
\emph{constructively}.  This is not entirely obvious: every textbook 
\emph{specification} of a stack contains axioms which are not functional.
It is easy to forget that a stack is an abstract data type, and so should
be treated as an abstract theory.  But, unfortunately, the abstract theories
of classical abstract data types have yet to be classified into an
organized whole, as have the abstract theories of mathematics.  We are 
working on this.

The kind of parametricity we offer is essentially that of
ML modules (see |List| in section~\ref{sec:concrete}).  This is suboptimal,
as we know that |List| could be made ``parametrically polymorphic''.  For
concrete theories, this makes no effective difference, but we are 
nevertheless unhappy with this.

\section{Applied Universal Algebra}

Universal Algebra is the study of algebraic structures themselves, rather
than models of algebraic structures.  In other words, rather than
studying groups, it is the \emph{theory} of groups which is studied.
Seen another way, it is the study of \emph{presentations of theories},
which is exactly what we have been dealing with in the previous two 
experiments.

What universal algebra brings is a uniform view of these, as well as
a number of \emph{constructions}.  Of course, category theory does the
same in a more general setting.  But for our purposes, it is the more
``concrete'' constructions of universal algebra which more readily bears fruit.

\subsection{The ideas}

We know that some constructions apply ``uniformly'' to most algebraic
structures.  For example, for single-sorted algebras, there is a uniform
notion of \emph{homomorphism} between them.  Thus, rather than trying to
have a human write what a $T$ homomorphism is for $>200$ theories $T$,
we hoped that we could automatically derive this from a presentation of $T$.
Similarly, we should be able to derive some notion of a sub-$T$-theory,
direct product, etc.

Furthermore, as we are in a setting where we have access to \emph{syntax}
as well as semantics, we can hope to derive the \emph{language} of a
theory automatically.  We wanted to explore if this was in fact feasible,
as well as see what other constructions we could automate.

\subsection{The experiment}

We chose to implement the following constructions:
\begin{enumerate}
  \item The construction of a type whose values represent the models
    of an arbitrary input theory.  As is done elsewhere, these are
    encoded as dependently-typed records (i.e. telescopes).

  \item The construction of the ``term algebra'' of a theory,
    as an inductive type.

  \item Automatically defining the concept of \emph{homomorphism}
    of an arbitrary input theory.

  \item Automatically defining the concept of \emph{substructure} 
    of an arbitrary input theory.

\end{enumerate}

\noindent Examples will illustrate these ideas better than formal definitions.
The declaration
\begin{lstlisting}
 type semigroup = TypeFrom(Semigroup)
\end{lstlisting}
\noindent in the context of a Theory means (i.e. is expanded to)
\begin{lstlisting}
 type semigroup = {U:type,*:(U,U)->U,
  associative_*:ProofOf(forall x,y,z : U.  (x * y) * z = x * (y * z))}
\end{lstlisting}
\noindent where |Semigroup| was defined in section~\ref{sec:abstract}.

Obtaining the \emph{term algebra} of a theory is just as simple:
\begin{lstlisting}
MonoidTerm := Theory { type MTerm = &Monoid }
\end{lstlisting}
\noindent denotes the inductive term
\begin{lstlisting}
MonoidTerm := Theory { 
  type MTerm = data X . 
    #e : X | 
    #* : (X, X) -> X
} 
\end{lstlisting}
\noindent |MTerm| is then exactly the set of free terms over the language of
Monoids.  We can then see associativity as an equation between two values
of |MTerm|.

We can continue in the same way for homomorphism.  We have that
\begin{lstlisting}
SemigroupH := Homomorphism(Semigroup)
\end{lstlisting}
\noindent means (expands to)
\begin{lstlisting}
SemigroupH := Theory {
  type SemiGroupType = TypeFrom(SemiGroup);
  A, B : SemiGroupType;
  f : A.U -> B.U;  
  axiom pres_*: forall x, y:A.U . f(x A.* y) = f(x) B.* f(y);    
}
\end{lstlisting}
\noindent  In other words, as expected, given two Semigroups, a homomorphism
is a function between their carrier sets which preserves multiplication.
In general, it is a function between carrier sets which preserves all 
operations, including nullary operations, aka constants.

Lastly, a substructure is one where a subset of the carrier set of a 
structure itself carries the structure.  For example,
\begin{lstlisting}
SubSemigroup := Substructure(Semigroup)
\end{lstlisting}
\noindent means (expands to)
\begin{lstlisting}
SubSemigroup := Theory {
  type SemiGroupType = TypeFrom(SemiGroup);
  A : SemiGroupType;
  V : type;
  axiom V <: A.U;
  axiom pres_*: forall x, y:V . defined-in(x A.* y, V)
}
\end{lstlisting}
\noindent where |<:| denotes subtyping and |defined-in| is a 
\emph{definedness} predicate, coming from the underlying logic.
When |A.U| (and thus |V|) is a set, this is just set membership.

\subsection{The method}

Implementing each of these transformations turns out to be quite
straightforward.  Each turns out to be a simple traversal of the
structure of a theory which maps each component in a precise manner
to the target.  The only difficulty is actually to decide on a
what form each concept (homomorphism, substructure, etc) should take.
Once this choice is made, the examples above give sufficient information
to extrapolate the implementation.  Given general enough 
traversal combinators, these are all less than $10$ lines of
Objective Caml code to implement (the combinators are O'Caml versions
of the Haskell package Multiplate~\cite{DBLP:journals/corr/abs-1103-2841}).

\subsection{The results}

The efficiency savings are quite significant: we can automatically obtain
definitions for the above $4$ concepts for all of our theories, $>200$ of
them, at a single stroke.  Furthermore, if we decide to make a change to
the details of how we want (say) sub-structures handled, we only have to
change our generator.  When we define new structures, we don't have to worry
about defining homomorphisms, sub-structure, etc for them, these are all
automatically derived for us.

Our method does highlight something frequently encountered in a
formalization context: textbook presentations of certain concepts 
frequently omit ``obvious'' axioms.  For example, we correctly generate
\begin{lstlisting}
axiom : forall x, y : V . f(h + h) = h(x) +' h(y);
axiom : forall a : F . forall x : V . h(a*x) = a*h(x);
axiom : forall x : V . h(-x) = -h(x);
axiom : h(0) = O';
axiom : h(1) = 1';
\end{lstlisting} 
\noindent for specifying a homomorphism of $F$-vector spaces, while 
textbooks will all too often only mention the first two axioms.

\section{Work of Influence}

The amount of related work is enormous, and even properly reviewing the work
which has had a clearly identifiable influence on us would more than double
the length of this paper.  We will restrict ourselves to highlighting a few
items which have had a significant impact on our work.

Some influences are obvious: the first author has learned heavily from the
successes and failures of Maple, while IMPS~\cite{FarmerEtAl93} served the
same purpose for the second author.  Further afield, the work of Douglas Smith%
\cite{Smith93,Smith99} on specification morphisms and the use of categorical
constructions in specifications should be clearly visible.

We have looked at quite a few libraries of mathematics, and we should in 
particular mention those of CASL~\cite{CoFI:2004:CASL-RM} and
Axiom~\cite{Jenks:1992:ASC} as well as the Wikipedia page~\cite{wiki:alg}
as sources of inspiration.  The wikipedia page gave us the right scope to
aim for, and CASL and Axiom's libraries, while nice, also convinced us that
too na\"{\i}ve an approach would take way too much human effort to achieve
our goals.

Last but not least, Parnas' ideas on modularization and information hiding%
\cite{journals/cacm/Parnas72a} and Dijkstra's on separation of
concerns~\cite{EWD:EWD447pub} are pervasive to our approach.  The root of
our use of \emph{generative} techniques as applied to mathematical
software~\cite{CaretteKiselyov11} actually finds its roots in Parnas' ideas
on Program Families~\cite{journals/tse/Parnas76}.

\section{Conclusion}

We knew that there was a lot of structure present in the theories of
mathematics.  But the principal lesson learned from this work is that
``a lot'' is a serious understatement.  More significantly, this structure can
be directly leveraged for the practical purposes of building a large library
of mathematics at a reasonable cost of human effort.  The resulting 
``source code'' is both readable and rather small in size, even though the
information it captures, in expanded form, is dauntingly large.

We also learned that there is also some ``higher order'' structure in these
theories: we can see duplication in our current library because we are
``replaying'' the same constructions on top of different starting theories.
We are currently working on capturing this structure present at the level
of \emph{graphs of theories}.  We actually stopped adding new theories to
our library as we found that a large number of them should have already been
present if we had features for working with graphs of theory constructions.

We also learned that there is still a fair amount of classification work
which needs to be done on the ``theories of computer science''.  We have 
gotten frequent glimpses of structure as rich as that known in mathematics,
but have not found these ideas in the literature.

Some theories are partly concrete and partly abstract: polynomials over an
arbitrary Ring are probably the best known example.  More subtly, 
polynomials are sometimes treated as syntactic entities, while other times
they are treated more semantically (if your representation for polynomials
$R[x]$ contains $x$, it is syntactic).  We are still perplexed by this.

\bibliography{mathscheme}
\bibliographystyle{plain}

\end{document}